\documentclass[twocolumn]{pasj00}
\usepackage[dvips]{color}

\begin{document}
\SetRunningHead{N. Narita et al.}{Direct Imaging of HAT-P-7}
\Received{2010/04/01}
\Accepted{2010/04/14}%{yyyy/mm/dd}

\title{Search for Outer Massive Bodies around Transiting Planetary Systems:
Candidates of Faint Stellar Companions around HAT-P-7$^*$}

%%% begin:list of authors

\author{
Norio \textsc{Narita},\altaffilmark{1,2}
Tomoyuki \textsc{Kudo},\altaffilmark{1}
Carolina \textsc{Bergfors},\altaffilmark{3}
Makiko \textsc{Nagasawa},\altaffilmark{2,4}
Christian \textsc{Thalmann},\altaffilmark{3}\\
Bun'ei \textsc{Sato},\altaffilmark{4}
Ryuji \textsc{Suzuki},\altaffilmark{5}
Ryo \textsc{Kandori},\altaffilmark{1}
Markus \textsc{Janson},\altaffilmark{6}
Miwa \textsc{Goto},\altaffilmark{3}
Wolfgang \textsc{Brandner},\altaffilmark{3}\\
Shigeru \textsc{Ida},\altaffilmark{2,7}
Lyu \textsc{Abe},\altaffilmark{8}
Joseph \textsc{Carson},\altaffilmark{3,9}
Sebastian E. \textsc{Egner},\altaffilmark{5}
Markus \textsc{Feldt},\altaffilmark{3}
Taras \textsc{Golota},\altaffilmark{5}\\
Olivier \textsc{Guyon},\altaffilmark{5,10}
Jun \textsc{Hashimoto},\altaffilmark{1}
Yutaka \textsc{Hayano},\altaffilmark{5}
Masahiko \textsc{Hayashi},\altaffilmark{5}
Saeko S. \textsc{Hayashi},\altaffilmark{5}\\
Thomas \textsc{Henning},\altaffilmark{3}
Klaus W. \textsc{Hodapp},\altaffilmark{11}
Miki \textsc{Ishii},\altaffilmark{5}
Gillian R. \textsc{Knapp},\altaffilmark{12}
Nobuhiko \textsc{Kusakabe},\altaffilmark{1}\\
Masayuki \textsc{Kuzuhara},\altaffilmark{1,13}
Taro \textsc{Matsuo},\altaffilmark{14}
Michael W. \textsc{McElwain},\altaffilmark{12}
Shoken \textsc{Miyama},\altaffilmark{1}
Jun-Ichi \textsc{Morino},\altaffilmark{1}\\
Amaya \textsc{Moro-Martin},\altaffilmark{15}
Tetsuo \textsc{Nishimura},\altaffilmark{5}
Tae-Soo \textsc{Pyo},\altaffilmark{5}
Eugene \textsc{Serabyn},\altaffilmark{14}
Takuya \textsc{Suenaga},\altaffilmark{1,16}\\
Hiroshi \textsc{Suto},\altaffilmark{1}
Yasuhiro \textsc{Takahashi},\altaffilmark{1,16}
Michihiro \textsc{Takami},\altaffilmark{17}
Naruhisa \textsc{Takato},\altaffilmark{5}
Hiroshi \textsc{Terada},\altaffilmark{5}\\
Daigo \textsc{Tomono},\altaffilmark{5}
Edwin L. \textsc{Turner},\altaffilmark{12,18}
Makoto \textsc{Watanabe},\altaffilmark{19}
Toru \textsc{Yamada},\altaffilmark{20}
Hideki \textsc{Takami},\altaffilmark{5}\\
Tomonori \textsc{Usuda},\altaffilmark{5} and
Motohide \textsc{Tamura}\altaffilmark{1}
}

\altaffiltext{1}{
National Astronomical Observatory of Japan, 2-21-1 Osawa,
Mitaka, Tokyo, 181-8588, Japan
}

\altaffiltext{2}{
Kavli Institute for Theoretical Physics, UCSB, Santa Barbara,
CA 93106-4030, USA
}

\altaffiltext{3}{
Max Planck Institute for Astronomy, 
K\"{o}nigstuhl 17, 69117, Heidelberg, Germany
}

\altaffiltext{4}{
Global Edge Institute, Tokyo Institute of Technology,
2-12-1 Ookayama, Meguro-ku, Tokyo, 152-8550, Japan
}

\altaffiltext{5}{
Subaru Telescope, 650 North A'ohoku Place, Hilo, HI 96720, USA
}

\altaffiltext{6}{
Department of Astronomy, University of Toronto, 50 St George St,
Toronto, ON M5S 3H4, Canada
}

\altaffiltext{7}{
Tokyo Institute of Technology, Ookayama, Meguro-ku,
Tokyo 152-8551, Japan
}

\altaffiltext{8}{
Laboratoire Hippolyte Fizeau, UMR6525, Universite de Nice Sophia-Antipolis,
28, avenue Valrose, 06108 Nice Cedex 02, France
}

\altaffiltext{9}{
Department of Physics and Astronomy,
College of Charleston, 58 Coming St., Charleston, SC 29424, USA
}

\altaffiltext{10}{
Steward Observatory, University of Arizona, Tucson, AZ 85719, USA
}

\altaffiltext{11}{
Institute for Astronomy, University of Hawaii,
640 N. Aohoku Place, Hilo, HI 96720, USA
}

\altaffiltext{12}{
Department of Astrophysical Sciences, Princeton University,
Peyton Hall, Ivy Lane, Princeton, NJ 08544, USA
}

\altaffiltext{13}{
University of Tokyo,
7-3-1 Hongo, Tokyo, 113-0033, Japan
}

\altaffiltext{14}{
Jet Propulsion Laboratory, 4800 Oak Grove Drive, MS 171-113,
California Institute of Technology, Pasadena, CA 91109, USA
}

\altaffiltext{15}{
Department of Astrophysics, CAB - CSIC/INTA,
28850 Torrej{\'o}n de Ardoz, Madrid, Spain
}

\altaffiltext{16}{
The Graduate University for Advanced Studies,
2-21-1 Osawa, Mitaka, Tokyo 181-8588, Japan
}

\altaffiltext{17}{
Institute of Astronomy and Astrophysics, Academia Sinica,
P.O. Box 23-141, Taipei 10617, Taiwan
}

\altaffiltext{18}{
Institute for the Physics and Mathematics of the Universe,
The University of Tokyo, Kashiwa 277-8568, Japan
}

\altaffiltext{19}{
Department of Cosmosciences, Hokkaido University,
Sapporo 060-0810, Japan
}

\altaffiltext{20}{
Astronomical Institute, Tohoku University,
Aoba, Sendai 980-8578, Japan
}
\email{norio.narita@nao.ac.jp}

%% `\KeyWords{}' always has to be placed before `\maketitle'.
%Do NOT move this preamble from here!

\maketitle

\begin{abstract}
We present results of direct imaging observations for HAT-P-7
taken with the Subaru HiCIAO and the Calar Alto AstraLux.
Since the close-in transiting planet HAT-P-7b was reported to have
a highly tilted orbit, massive bodies such as giant planets,
brown dwarfs, or a binary star are expected to exist
in the outer region of this system.
We show that there are indeed two candidates for distant
faint stellar companions around HAT-P-7.
We discuss possible roles played by such companions
on the orbital evolution of HAT-P-7b. We conclude
that as there is a third body in the system as reported
by Winn et al. (2009, ApJL, 763, L99), the Kozai migration
is less likely while planet-planet scattering is possible.
\end{abstract}
\footnotetext[*]{Based on data collected at Subaru Telescope,
which is operated by the National Astronomical Observatory of Japan.}

\section{Introduction}

The discovery of over 400 extrasolar planets and the diversity
of their orbital distributions dramatically changed our perception
of planetary systems in the last 15 years.
Especially, the existence of exoplanets in very close-in
or eccentric orbits stimulated theorists to develop various models
for planetary migration during the epoch of planet formation.
To explain the orbital distribution of known exoplanets,
a number of planetary migration models have been proposed,
including disk-planet interaction models
(i.e., Type I and Type II migration models; e.g.,
\cite{1985prpl.conf..981L, 1996Natur.380..606L, 2002A&A...385..647D,
2004ApJ...616..567I}),
planet-planet scattering models considering
gravitational interaction among multiple giant planets
(i.e., jumping Jupiter models;
e.g., \cite{1996Sci...274..954R, 2002Icar..156..570M, 
2008ApJ...678..498N, 2008ApJ...686..580C}),  or
Kozai migration models considering perturbation by
a distant massive companion and coinstantaneous tidal evolution
(e.g., \cite{2003ApJ...589..605W, 2005ApJ...627.1001T, 
2007ApJ...669.1298F, 2007ApJ...670..820W}).

These planetary migration models can now be tested
by observations of the Rossiter-McLaughlin effect
(hereafter the RM effect: \cite{1924ApJ....60...15R,
1924ApJ....60...22M}) for transiting planetary systems,
which is an anomalous shift in observed radial velocities
due to the occultation of a rotating star.
Measurements of the RM effect enable us to estimate
the sky projection angle of the planetary orbital axis
relative to the stellar spin axis
(i.e., the spin-orbit alignment angle;
\cite{2005ApJ...622.1118O, 2007ApJ...655..550G, 2010ApJ...709..458H}).
The information of the spin-orbit alignment angle is very useful
to differentiate the planet-planet scattering models and the Kozai
migration models from the disk-planet interaction models,
because the former models predict a wider range of spin-orbit
alignment angles for migrated planets,
while the latter models predict that
migrated planets would have only small spin-orbit alignment angles.
Thus an observation of a highly tilted orbit
for a specific planet is strong evidence of
a planet-planet scattering process or perturbation
by an outer companion during its migration history.

Indeed, very recently several transiting exoplanets
have been reported to show such highly tilted orbits
via measurements of the RM effect.
These observations lead to an interesting prediction:
around these spin-orbit misaligned exoplanets,
other massive bodies (e.g., massive planets, brown dwarfs,
or a low-mass stellar companion) should be present.
In addition, since migration mechanisms for such exoplanets
cannot be distinguished by the spin-orbit alignment angles alone,
direct imaging of such massive bodies gives us important
additional information to distinguish between the two migration
mechanisms of highly tilted orbit planets.
Motivated by these facts, we have initiated to search for such
outer massive bodies around known transiting planetary systems
with the Subaru HiCIAO (High Contrast Instrument for the Subaru
next generation Adaptive Optics; \cite{2006SPIE.6269E..28T,
2008SPIE.7014E..42H}; Suzuki et al. 2010, in prep.),
as part of the SEEDS project
(Strategic Explorations of Exoplanets and Disks with Subaru,
PI: Motohide Tamura).
The Subaru HiCIAO is a powerful instrument to search for outer
faint bodies around stars, proven by the detection of a massive planet
or a brown dwarf around GJ~758 \citep{2009ApJ...707L.123T}.

In this paper, we targeted the transiting planetary system HAT-P-7,
which was reported to have a planet (HAT-P-7b) on an orbit highly
inclined relative to the stellar equatorial plane
(\cite{2009PASJ...61L..35N}; hereafter NSH09,
\cite{2009ApJ...703L..99W}; hereafter WJA09).
Consequently, we report two candidates of faint stellar
companions to HAT-P-7 based on the Subaru HiCIAO and
the Calar Alto AstraLux data.
Although our data alone cannot distinguish whether or not the candidate
companion stars are physically associated with HAT-P-7 at this point,
the findings are useful to constrain the mechanism of
planetary migration in this system.
We summarize the properties of our target in section~2,
and report our observations, analyses, and results in section~3.
We present theoretical discussions of the migration mechanism
of HAT-P-7b in section~4.
Finally, section~5 summarizes the findings in this paper.

%%%%%%%%%%%%%%%%%%%%%%%%%%%%%%%%%%%%%%%%%%%%%%%%%%%%%%%%%%%%%%%%%%%%%%
\begin{table}[t]
\caption{Summary of stellar and planetary parameters.}
\begin{center}
\begin{tabular}{l|ccc}
\hline
Parameter & Value & Error & Source \\
\hline
Star & & & \\
$M_s$ [$M_{\odot}$] 
& $1.520$ & 0.036 & CKB10 \\
$R_s$ [$R_{\odot}$]
& $1.991$ & 0.018 & CKB10 \\
Age [Gyr]
& $2.14$ & 0.26 & CKB10 \\
Distance [pc]
& $320$ & $^{+50}_{-40}$ & PBT08 \\
app. $H$ mag.
& 9.344 & 0.029 & \citet{2003tmc..book.....C} \\
\hline
Planet & & & \\
$M_p$ [$M_{Jup}$] 
& $1.82$ & 0.03 & WOS10 \\
$R_p$ [$R_{Jup}$]
& $1.50$ & 0.02 & WOS10 \\
$i$ [$^{\circ}$]
& $83.1$ & 0.5 & WOS10 \\
$a$ [AU]
& 0.0386 & 0.0001 & WOS10 \\
$P$ [days]
& $2.204733$ & 0.000010 & WOS10 \\
\hline
\multicolumn{4}{l}{\hbox to 0pt{\parbox{80mm}{\footnotesize
}\hss}}
\end{tabular}
\end{center}
\end{table}
%%%%%%%%%%%%%%%%%%%%%%%%%%%%%%%%%%%%%%%%%%%%%%%%%%%%%%%%%%%%%%%%%%%%%%

\section{Target Properties}

HAT-P-7 (also known as Kepler-2) is an F8\footnote{There
is a slight uncertainty in the spectral subclass of HAT-P-7, and
the uncertainty would have a slight effect on apparent $i'$ and $z'$
band magnitudes quoted in table~2. However, subsequent discussions
and conclusion of this paper would remain unchanged.} star,
hosting a very hot Jupiter HAT-P-7b
(\cite{2008ApJ...680.1450P}; hereafter PBT08).
The stellar distance was estimated as $320^{+50}_{-40}$ pc (PBT08).
According to 2MASS catalog \citep{2003tmc..book.....C},
the $H$ band magnitude of HAT-P-7 is $9.344 \pm 0.029$.
This star is in the field of view of the NASA Kepler mission
\citep{2009Sci...325..709B},
and detailed stellar parameters were reported
through a Kepler asteroseismology study as follows;
the mass $M_s = 1.520 \pm 0.036$ $M_{\odot}$,
the radius $R_s = 1.991 \pm 0.018$ $R_{\odot}$, and
the age $2.14 \pm 0.26$ Gyr (\cite{2010arXiv1001.0032C}; hereafter CKB10).
The planet HAT-P-7b has a very close-in orbit
with the orbital period of $2.204733 \pm 0.000010$ days and
the semi-major axis of $a = 0.0386 \pm 0.0001$ AU
(\cite{2010arXiv1001.0413W}; hereafter WOS10).
The mass, radius, and orbital inclination of HAT-P-7b are
$1.82 \pm 0.03 M_{Jup}$,
$1.50 \pm 0.02 R_{Jup}$, and
$83.1^{\circ} \pm 0.5^{\circ}$, respectively (WOS10).
These properties are summarized in table~1.

The orbit of HAT-P-7b does not have a significant eccentricity
(PBT08, NSH09, WJA09, WOS10).
Nevertheless, NSH09 and WJA09 found that the planet has an
extremely tilted orbit relative to the stellar rotation axis.
In addition, WJA09 reported an additional long term
RV trend, implying a third body in the planetary system.
Thus this system is a very fascinating target
of direct imaging to search for outer massive bodies.

%%%%%%%%%%%%%%%%%%%%%%%%%%%%%%%%%%%%%%%%%%%%%%%%%%%%%%%%%%%%%%%%%%%%%%
\begin{figure*}[phtb]
 \begin{center}
  \FigureFile(160mm,160mm){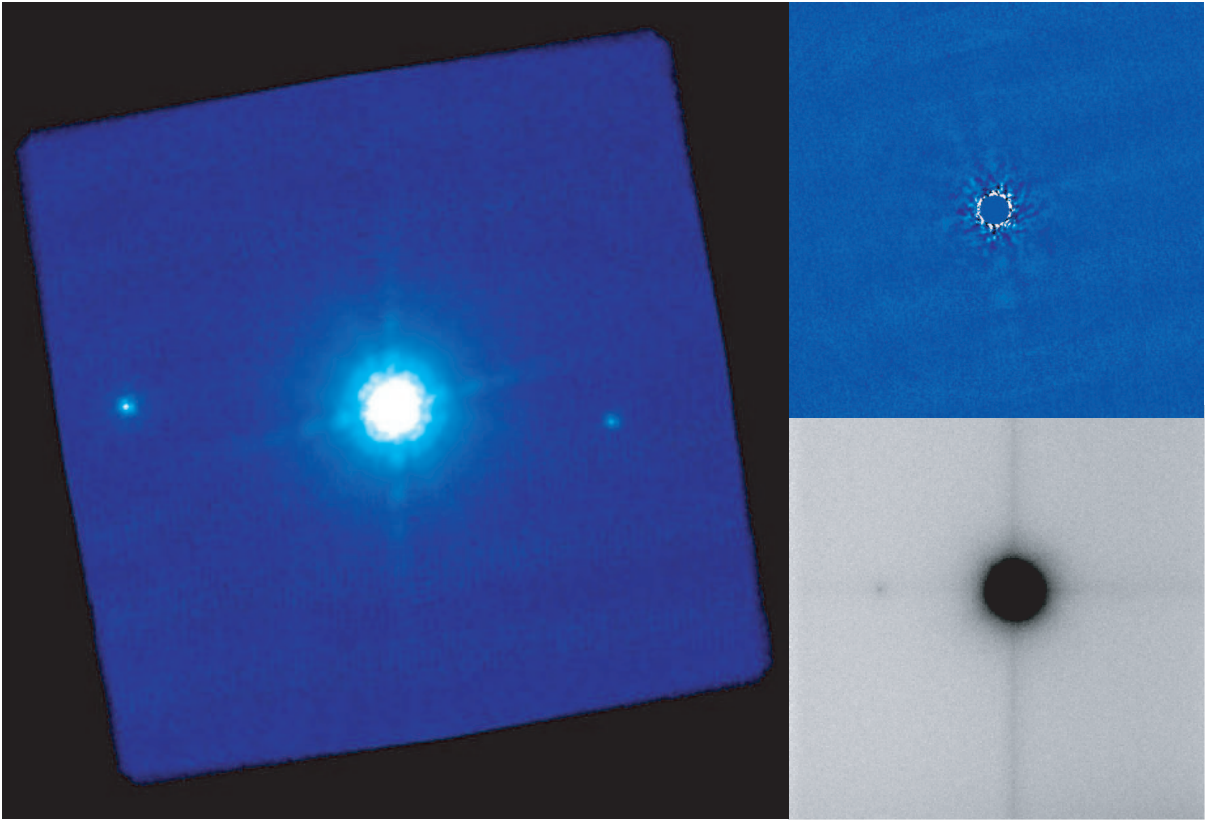}
 \end{center}
  \caption{Left: A median combined ADI image of HAT-P-7 in $H$ band
  taken with the Subaru HiCIAO on UT~2009~August~6.
  The field of view is 12$\times$12 arcsec.
  Upper right: A ADI/LOCI reduced Subaru image of the inner region
  around HAT-P-7. The field of view is 6$\times$6 arcsec.
  Lower right: AstraLux $z'$ band image of HAT-P-7 and the eastern companion
  candidate. The field of view is 12$\times$12 arcsec.
  North is up and east is left for all panels.
  }
\end{figure*}
%%%%%%%%%%%%%%%%%%%%%%%%%%%%%%%%%%%%%%%%%%%%%%%%%%%%%%%%%%%%%%%%%%%%%%

\section{Analyses}

\subsection{Subaru / HiCIAO}

We first observed HAT-P-7 in the $H$ band with the HiCIAO
combined with the AO188 (188-element curvature sensor
adaptive optics system; \cite{2008SPIE.7015E..25H}),
mounted on the 8.2 m Subaru Telescope on UT~2009~August~6.
The field of view was 20$\times$20 arcsec, and
typical natural seeing was about 0.5 arcsec on that night.
We used the target star itself as natural guiding for AO188.
We took 30 object frames with 19.50 s exposures
(i.e., total exposure time was 9.75 min).
The observations were conducted in pupil tracking mode to use
the angular differential imaging (ADI: \cite{2006ApJ...641..556M})
technique.
The gain of the detector was 1.66~e$^-$ per ADU,
readout noise was 15~e$^-$, and zero point magnitude
(the magnitude of an object that would yield 1 ADU per s)
of the image was $H=24.654\pm0.036$ mag.
Our Subaru HiCIAO reduction procedures were as follows.
We first removed a characteristic stripe bias pattern
arising from the Subaru HiCIAO detector.
Then flat fielding was done using dome flat frames,
and bad/hot pixels were removed.
Note that we did not subtract dark frames since
dark current of Subaru HiCIAO is sufficiently low for
the exposure time.
We corrected distortion of the field of view
by comparison of an M15 image taken by Subaru HiCIAO on
UT 2009 August 5 with that taken by HST/ACS.
The measured pixel scale of Subaru HiCIAO was $9.44\pm0.10$ mas per pixel.
We shifted frames to match stellar centroid and
used the ADI technique to combine the object frames.
Field rotation during our exposures was $7.023^{\circ} \pm 0.007^{\circ}$.

The median combined object frame is shown in the left panel of figure~1
(north is up and east is left, and the field of view is
12$\times$12 arcsec, as a subset of the full 20$\times$20 arcsec frame).
Two faint sources at about 3 arcsec away were clearly detected.
The FWHM of the PSF was 6.1 pixel (0.058 arcsec).
We determined positions of the candidate companion stars
using the \textit{imexam} task in IRAF\footnote{The Image
  Reduction and Analysis Facility (IRAF) is distributed by the U.S.\
  National Optical Astronomy Observatories, which are operated by the
  Association of Universities for Research in Astronomy, Inc., under
  cooperative agreement with the National Science Foundation.}
and conducted aperture photometry using the \textit{phot} task.
We used the photometric standard star FS151 ($H=11.946\pm0.008$ mag)
to determine apparent $H$ band magnitudes of the candidate companion stars,
since HAT-P-7 in the image was saturated.
The apparent $H$ band magnitude, separation angle, and position angle of
each companion star from HAT-P-7 are listed
in the left column of table~2.
Combining our result for the separation angle and the PBT08 result for
the stellar distance, we estimate that projected separation distances
of these candidate companion stars from HAT-P-7 are about 1000 AU.

In addition, we used the locally optimized combination of images
algorithm (LOCI, \cite{2007ApJ...660..770L}) to maximize the
efficiency of the ADI technique and to search for fainter objects
in the inner region around HAT-P-7.
The upper right panel of figure~1 shows the LOCI reduced image around
HAT-P-7 (north is up and east is left, and the field of view is
6$\times$6 arcsec centered on HAT-P-7, as a subset of the full
20$\times$20 arcsec frame).
Clearly, the bright halo in the inner region seen in the left panel
is significantly suppressed.
The upper panel of figure~2 plots 5$\sigma$ contrast ratio around HAT-P-7
achieved by the Subaru HiCIAO observation, and the lower panel of figure~2
shows corresponding detectable mass of companions,
assuming the age of 2.14 Gyr and the COND model by \citet{2003A&A...402..701B}.
We achieved post-LOCI $5\sigma$ contrast sensitivity for $H$ band
of $\sim7\times10^{-4}$ at 0.3 arcsec, $\sim2\times10^{-4}$ at 0.5 arcsec,
and $\sim6\times10^{-5}$ at 1.0 arcsec,
corresponding to $\sim110$ $M_{Jup}$, $\sim80$ $M_{Jup}$,
and $\sim70$ $M_{Jup}$ companions, respectively.
As a result, we exclude the presence of a stellar companion
(more massive than 80 $M_{Jup}$ within the 6$\times$6 arcsec
field of view) separated 0.5 arcsec or farther
at the $5\sigma$ level.
At this point, we have not yet put a stringent constraint
on inner massive bodies.
For instance, an M star ($\sim 100 M_{Jup}$) is not ruled out within
0.3 arcsec ($\sim 100$ AU).

%%%%%%%%%%%%%%%%%%%%%%%%%%%%%%%%%%%%%%%%%%%%%%%%%%%%%%%%%%%%%%%%%%%%%%
\begin{figure}[pthb]
 \begin{center}
  \FigureFile(80mm,80mm){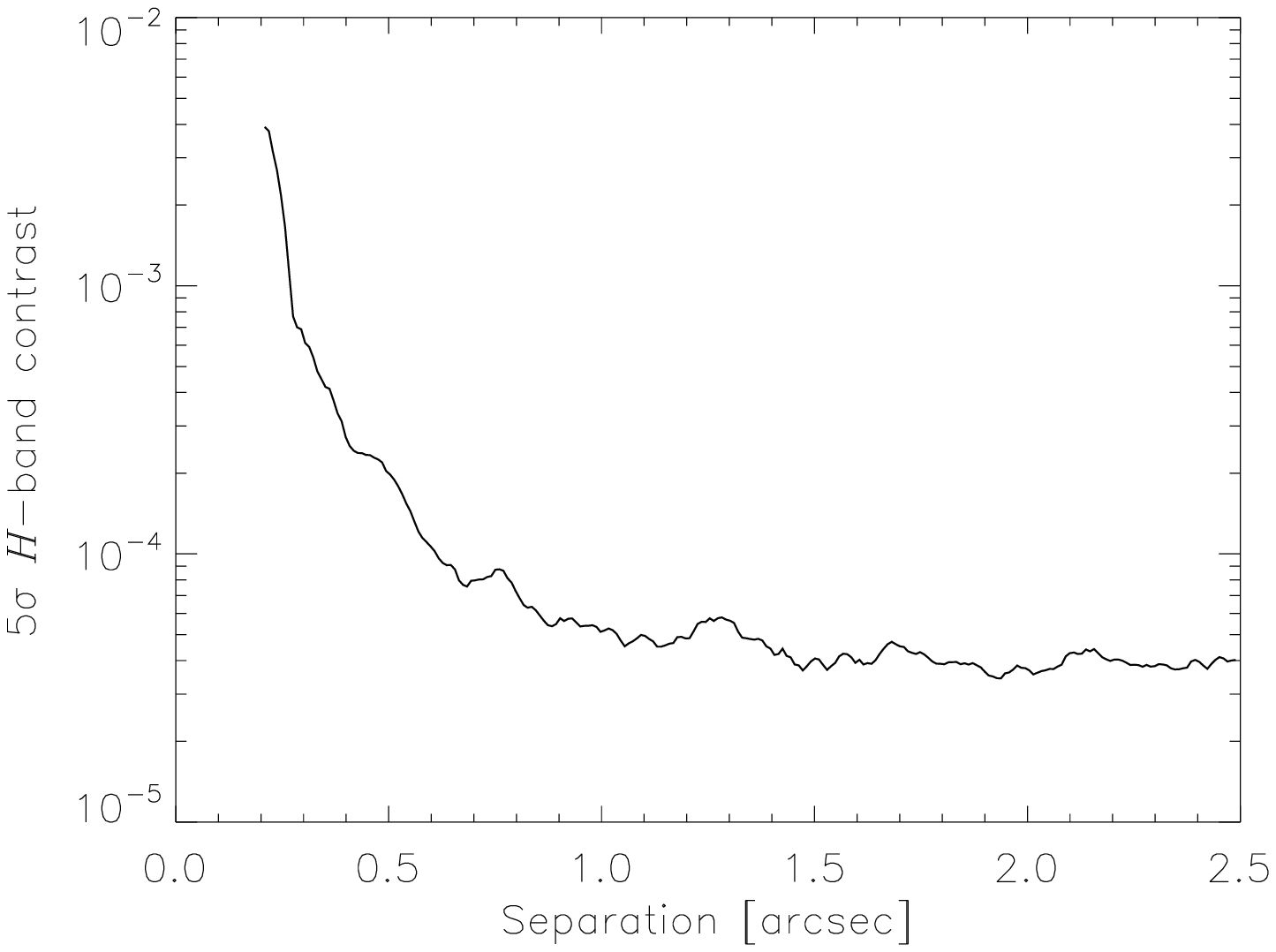}
  \FigureFile(80mm,80mm){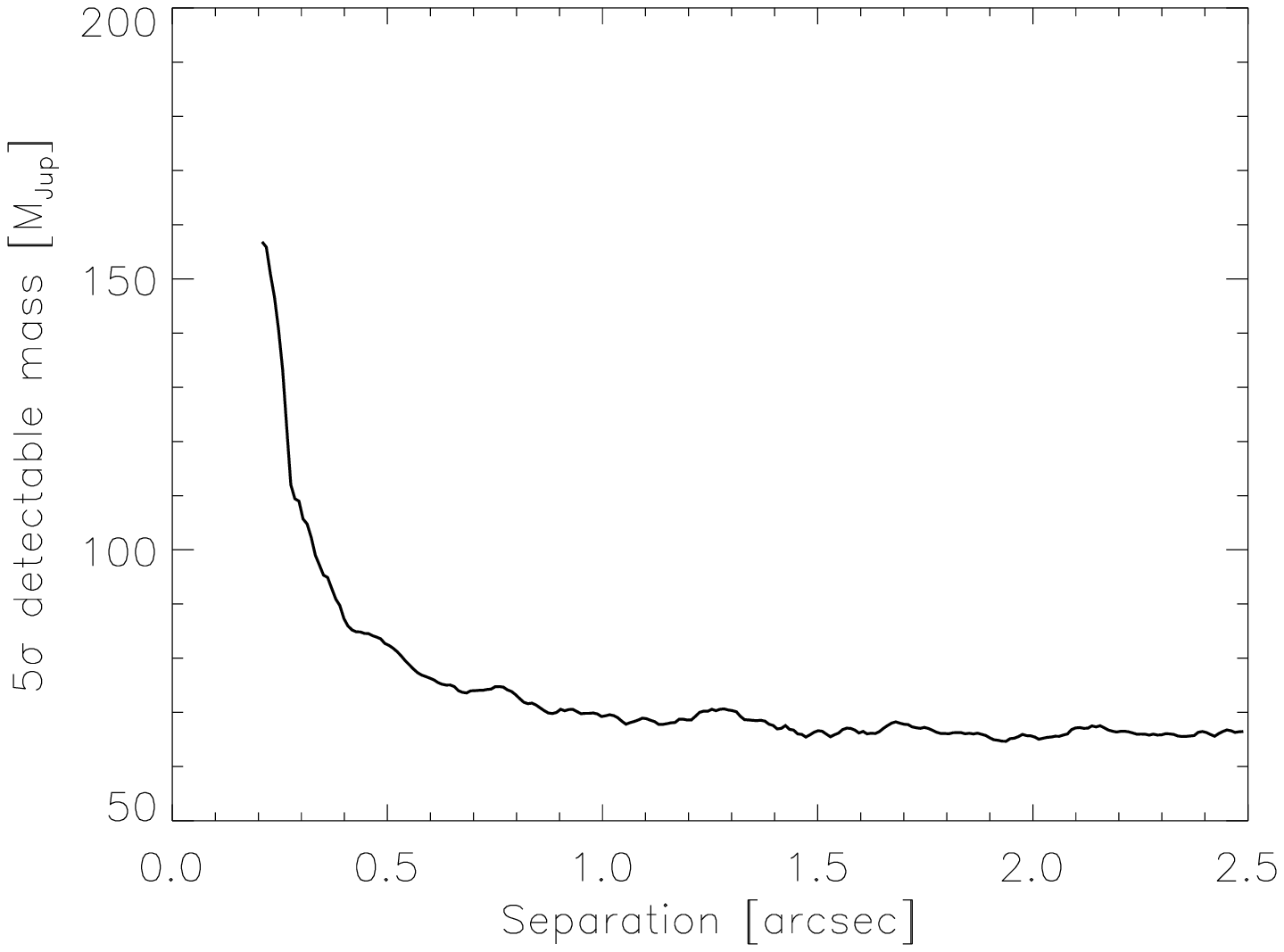}
 \end{center}
  \caption{Upper: ADI/LOCI 5$\sigma$ contrast ratio around HAT-P-7
  achieved by the Subaru HiCIAO.
  Lower: Corresponding constraint on detectable mass of companions,
  assuming the age of 2.14 Gyr and the COND model by
  \citet{2003A&A...402..701B}.
  We achieved the contrast ratio of $\sim7\times10^{-4}$ (0.3 arcsec),
  $\sim2\times10^{-4}$ (0.5 arcsec), and $\sim6\times10^{-5}$ (1.0 arcsec),
  corresponding to the detectable mass of $\sim110$ $M_{Jup}$,
  $\sim80$ $M_{Jup}$, and $\sim70$ $M_{Jup}$, respectively.
  }
\end{figure}
%%%%%%%%%%%%%%%%%%%%%%%%%%%%%%%%%%%%%%%%%%%%%%%%%%%%%%%%%%%%%%%%%%%%%%

%%%%%%%%%%%%%%%%%%%%%%%%%%%%%%%%%%%%%%%%%%%%%%%%%%%%%%%%%%%%%%%%%%%%%%
\begin{figure}[pthb]
 \begin{center}
  \FigureFile(80mm,80mm){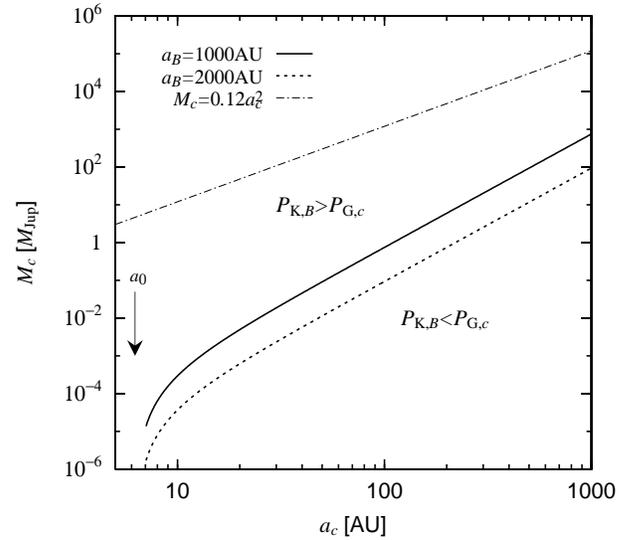}
 \end{center}
  \caption{Boundaries of the restricted area where HAT-P-7c cannot exist
  initially, since the timescale of orbital precession of HAT-P-7b caused
  by gravitational perturbation from HAT-P-7c ($P_{{\rm{G}},c}$) is shorter
  than that caused by the Kozai mechanisum due to the distant companion
  ($P_{{\rm{K}},B}$).
  The solid line is for $a_B=1000$ AU and the dotted line is for
  $a_B=2000$ AU. The dashed-dotted line indicates a relation for the
  semi-major axis and the mass of HAT-P-7c reported by Winn et al. (2009).
  $a_0$ indicates a roughly estimated position of the snow line.
  }
\end{figure}
%%%%%%%%%%%%%%%%%%%%%%%%%%%%%%%%%%%%%%%%%%%%%%%%%%%%%%%%%%%%%%%%%%%%%%

%%%%%%%%%%%%%%%%%%%%%%%%%%%%%%%%%%%%%%%%%%%%%%%%%%%%%%%%%%%%%%%%%%%%%%
\begin{table*}[t]
\caption{Magnitudes, positions, estimated$^\dagger$ spectral
type and masses of candidate companion stars.}
\begin{center}
\begin{tabular}{l|cccccc}
\hline
 &
\multicolumn{2}{c}{HiCIAO ($H$)} &
\multicolumn{2}{c}{AstraLux ($i'$)$^\ddagger$} &
\multicolumn{2}{c}{AstraLux ($z'$)$^\ddagger$} \\
 &
\multicolumn{2}{c}{2009 August 6} &
\multicolumn{2}{c}{2009 October 30} &
\multicolumn{2}{c}{2009 October 30} \\
Parameter & Value & Error & Value & Error & Value & Error \\
\hline
West (fainter) & & & & \\
apparent magnitude [mag] & 16.92 & 0.06 &
 $>$18.65 & --   & $>$18.55  & --  \\
separation angle [''] & 3.14 & 0.01 &
 -- & --   & --   & --  \\
position angle [$^{\circ}$] & 266.30 & 0.37 &
 -- & --   & --   & --  \\
Estimated Spectral Type and Mass [$M_\odot$]&
\multicolumn{6}{c}{M9V-L0V (0.078-0.088)} \\
\hline
East (brighter) & & & & \\
apparent magnitude [mag] & 15.12 & 0.04 &
 18.50 &  0.21  & 17.43   & 0.09  \\
separation angle [''] & 3.88 & 0.01 &
 -- & -- &  3.82  & 0.01 \\
position angle [$^{\circ}$] & 89.81 & 0.30 &
 -- & -- & 90.39 & 0.11 \\
Estimated Spectral Type and Mass [$M_\odot$]&
\multicolumn{6}{c}{M5V-M6V (0.17-0.20)} \\
\hline
\multicolumn{7}{l}{\hbox to 0pt{{\footnotesize
$^\dagger$Assuming that the candidate companions are main sequence
stars and at the same distance as HAT-P-7.}\hss}}\\
\multicolumn{7}{l}{\hbox to 0pt{{\footnotesize
$^\ddagger$Assuming that HAT-P-7 is an F8 star.}\hss}}
\end{tabular}
\end{center}
\end{table*}
%%%%%%%%%%%%%%%%%%%%%%%%%%%%%%%%%%%%%%%%%%%%%%%%%%%%%%%%%%%%%%%%%%%%%%

\subsection{Calar Alto / AstraLux Norte Lucky Imaging}

HAT-P-7 was observed in SDSS $i'$ and $z'$ filter with
the AstraLux Norte Lucky Imaging camera \citep{2008SPIE.7014E.138H}
at the 2.2 m telescope at Calar Alto on UT 2009 October 30.
The observations were part of a large-scale high resolution imaging search
for close stellar companions to all known exoplanet hosts brighter
than $i'=16$ mag (see \cite{2009A&A...498..567D} for the first results).
The survey employs the Lucky Imaging technique, which provides
almost diffraction limited images by shift-and-add drizzle combination
of only the best few percent of a series of $\sim$10000 very short
exposures ($\sim$10 ms), selected by the Strehl ratio.
For the photometric analysis of HAT-P-7 and its companion candidates,
the best 10\% of a total 20000 integrations of 15 ms exposure
time were used, yielding a total integration time of 30 s.
The combined $z'$ band image is presented in the lower right panel
of figure~1 (north is up and east is left, and the field of view is
12$\times$12 arcsec).

The IRAF \textit{phot} task was used for relative aperture photometry
of HAT-P-7 and the eastern candidate companion.
The western companion could not be seen in the AstraLux images.
Combining the relative photometry with the $JHK$ magnitudes of
HAT-P-7 \citep{2003tmc..book.....C} and the absolute magnitudes
of an F8 main sequence star \citep{2007AJ....134.2340K},
we derive the apparent $i'$ and $z'$ band magnitudes as
$i'=18.50\pm0.21$ mag and $z'=17.43\pm0.09$ mag for the eastern companion
candidate and upper limits of $i'>18.65$ mag and $z'>18.55$ mag
for the western companion candidate.
The astrometric calibration was achieved following the procedures
outlined by \citet{2008A&A...482..929K} based on observations of
stars with well defined astrometry in the Orion Nebula Cluster.
The derived image scale of AstraLux Norte for this observing
run was $23.43\pm0.06$ mas per pixel.
For the eastern companion candidate, we determined a separation
of $3.82\pm0.01$ arcsec and a position angle of $90.39\pm0.11$ deg.
These results are summarized in table~2.

\subsection{Combined Results}

We discovered two candidates of faint stellar companions
around HAT-P-7.
As for the eastern companion,
the $i'$-$z'$ colors for the eastern companion suggest a main sequence
spectral type of M5V-M6V, corresponding to 0.17-0.20$M_\odot$
\citep{2007AJ....134.2340K, 2007AJ....134.2398C}.
For this luminosity class, its apparent
brightness in $z'$ band yields as distance modulus of 10.03 mag for
spectral type M6V, corresponding to a distance of $\sim$300 pc.
It is in agreement with the distance estimate for HAT-P-7 of
$320^{+50}_{-40}$ pc \citep{2008ApJ...680.1450P}.
We note that the magnitude and color of the companion candidate
would also be explained by a background M5III-M6III star
at the distance of 126 kpc or more \citep{2007AJ....134.2398C}.
As for the western companion, the AstraLux Norte observations
only provide an upper limit on the $i'$ and $z'$ band brightness;
we can just derive a lower limit on its $i'$-$H$ and $z'$-$H$ colors.
Assuming that the western companion candidate is a main sequence star
associated with HAT-P-7, its $H$ band magnitude suggest M9V-L0V,
corresponding to a mass in the range 0.078-0.088$M_\odot$
(see \cite{2007AJ....134.2340K}).

We note that, however, it is unrealistic that both
stars are true companions of HAT-P-7, because such a wide
separation triple system is likely to be physically unstable
and it is known that triple or quadruple stars discovered so far
are hierarchial \citep{1991A&A...248..485D}.
High spectral resolution observations
for the candidate companion stars would be useful to
constrain their peculiar radial velocities
and thereby to discriminate their binarity.
In addition, follow-up direct imaging observations,
in the near future, will allow us to show via common proper motion
whether the objects are true companions of HAT-P-7, whose proper
motion is $18.60 \pm 3.25$ mas per year in the TYCHO reference catalogue
\citep{1998A&A...335L..65H}.
We note, however, that there are over 3$\sigma$ inconsistency
between the Subaru HiCIAO result and the Calar Alto AstraLux result
(epoch difference of less than 3 months)
for the separation angle of the eastern companion.
However, at this point we do not conclude that the eastern companion
is not associated with HAT-P-7, because the separation angle
might be affected by the errors of the pixel scale,
and also because the errors of the separation angle
might be underestimated due to systematic errors caused by
saturation or non-gaussian PSF shape.
Thus further follow-up observations would be needed to distinguish
whether the two stars are truly associated with HAT-P-7 or not.

\section{Discussion}

In this section, we discuss realizable migration mechnisms
of the highly tilted orbit planet HAT-P-7b.
First of all, as a simple case, if neither of the candidate companions
we discovered is physically associated with HAT-P-7, only planet-planet
scattering can explain the tilted orbit of HAT-P-7b.
We thus examine a Kozai migration scenario for HAT-P-7b first,
based on an assumption that either of the candidate companion stars
is a true binary of HAT-P-7.
We present conditions required for the Kozai migration
of HAT-P-7b with a binary companion in section 4.1.
We show a restricted area of a third body
for the Kozai migration in this system in section 4.2., and
describe the impact of the possible third body (HAT-P-7c)
reported by WJA09 in section 4.3.

\subsection{Required Condition for the Kozai Migration of HAT-P-7b}

According to \citet{1962AJ.....67..591K}, the angular momentum
\begin{equation}
L_Z \equiv \sqrt{G (M_p + M_s) a (1 - e^2)} \cos \Psi_m
\end{equation}
should be conserved in a planetary system with a binary star
during the Kozai mechanism,
where $G$ is the gravitational constant,
$M_p$ and $M_s$ are the mass of the planet and its host star,
$a$ and $e$ are the semi-major axis and the orbital eccentricity of
the planet, and finally $\Psi_m$ (domain: [0$^{\circ}$, 180$^{\circ}$])
is the mutual inclination between the orbital inclination of
the planet and the binary star.
In addition, given that the angular momentum is also conserved
during tidal orbital evolution, $a (1-e^2) \cos^2 \Psi_m$ should be
conserved through the Kozai migration.

Using the conservation relation above,
we constrain the initial mutual inclination to initiate
the Kozai migration in this system as follows.
First we assume that HAT-P-7b was born in the snow line
with the initial eccentricity $e_0$ and the initial
mutual inclination $\Psi_{m,0}$.
The distance of the snow line from the host star is roughly estimated
as $a_0 = 2.7 (M_s/M_\odot)^2 = 6.24$ AU (i.e. 20 mas).
We note that although the position of the snow line is somewhat uncertain,
this makes little impact on the following discussions.
Then using the conservation relation rewritten as
\begin{equation}
a_0 (1-e_0^2) \cos^2 \Psi_{m,0}
= a_n (1-e_n^2) \cos^2 \Psi_{m,n},
\end{equation}
where the indices $0$ and $n$ indicate values of
the initial state and those as of now,
we obtain
\begin{equation}
|\cos \Psi_{m,0}| \le
\sqrt{\frac{0.0386}{6.24} \frac{1}{1-e_0^2}} \,\, |\cos \Psi_{m,n}|,
\end{equation}
as a necessary initial condition for the orbit of HAT-P-7b.
If this condition is satisfied, the eccentricity would be excited
over the critical value $\sqrt{1-0.0386/6.24}=0.997$ and
the planet would initiate tidal evolution.

We note that the timescale of the Kozai migration under consideration
is approximated as \citep{2007ApJ...670..820W},
\begin{equation}
P_{\tiny{\textrm{K}}} \sim \frac{M_s}{M_B} \frac{P_B^2}{P_{0}} (1-e_B^2)^{3/2},
\end{equation}
where $M_B$, $P_B$, and $e_B$ are the mass, orbital period, and eccentricity
of the binary star, and $P_{0}$ is the orbital period of HAT-P-7b
at the initial position.
Assuming that $M_B = 0.20 M_{\odot}$ (as a typical mass
of M star) and $e_B = 0$, we obtain $P_{\tiny{\textrm{Kozai}}} \sim$ 300 Myr,
which is sufficiently short relative to
the age of this system ($\sim2$ Gyr).
In addition, the timescale of general relativity is estimated as
\citep{2003ApJ...589..605W},
\begin{equation}
P_{\tiny{\textrm{GR}}} \sim
\frac{2 \pi c^2 (1-e_0^2) a_0^{5/2}}{3 (GM_s)^{3/2}}
\sim 2 \textrm{Gyr},
\end{equation}
where $c$ is the speed of light.
Thus general relativity would not disturb the Kozai migration
of HAT-P-7b at an early stage.

Based on equation (3), if HAT-P-7b was born with $e_0 = 0$
and if $\Psi_{m,n}=0^{\circ}$, namely if HAT-P-7b
and the binary star are coplanar now,
the initial mutual inclination $\Psi_{m,0}$ needs to be within
$85.5^{\circ}-94.5^{\circ}$ to initiate the Kozai migration.
Even if we assume that the initial eccentricity is large
(e.g., $e_0=0.8$) and the current mutual inclination is zero,
$\Psi_{m,0}$ needs to be within $82.5^{\circ}-97.5^{\circ}$.
This is a very optimistic case; if the eccentricity is lower and
$\Psi_{m,n}$ is not zero, the required condition becomes more stringent.
These required conditions are very tight,
but still possible
(a few suggestive circumstellar disk observations of nonzero
$\Psi_{m,0}$ for young binary stars were reported; e.g.,
\cite{1998ApJ...505..358A,2005ApJ...628..832D,2009PASJ...61.1271H}).

\subsection{Restricted Area of a Third Body for
the Kozai Migration Scenario}

As discussed by \citet{2003ApJ...589..605W} for HD~80606b,
a hypothetical additional body HAT-P-7c in the HAT-P-7 system could
destroy the Kozai migration process \citep{1997AJ....113.1915I},
if the timescale of orbital precession of HAT-P-7b caused by
the gravitational perturbation from HAT-P-7c ($P_{{\rm{G}},c}$) is
shorter than that caused by the
Kozai mechanism due to the binary companion ($P_{{\rm{K}},B}$).
We calculated a conditional equation for a restricted area of
an outer third body at the initial stage as,
\begin{equation}
M_c > \frac{3}{2} M_s \frac{a_c^2 a}{a_B^3} \frac{1}{b_{3/2}^{(1)}},
\end{equation}
where $a_c$ and $M_c$ are the semi-major axis and mass of
the additional planet, $a_B$ is the semi-major axis of the binary star,
and $b_{3/2}^{(1)}$ is the Laplace coefficient
(see e.g., \cite{2000ssd..book.....M, 2003ApJ...589..605W}).

The boundary of the restricted area is plotted in figure~3 by
solid (for $a_B=1000$ AU) and dotted (for $a_B=2000$ AU) lines.
The horizontal axis and the vertical axis represent $a_c$ and $M_c$,
respectively.
More specifically, the upper region of the solid (dotted) line
is the restricted area where HAT-P-7c cannot exist initially
for the Kozai migration caused by the binary companion.
This constraint is very stringent, and even analogies of
Saturn ($a_c = 9.6$ AU, $M_c = 0.3 M_{Jup}$) or
Uranus ($a_c = 19.2$ AU, $M_c = 0.04 M_{Jup}$)
cannot exist.

\subsection{Impact of HAT-P-7c Reported by Winn et al. (2009)}

On the other hand, WJA09 reported that there is indeed a possible
third body HAT-P-7c in the HAT-P-7 system
(hereafter, just ``c'').
As a constraint on the mass and semi-major axis of
the additional body, WJA09 reported the following equation;
\begin{equation}
\frac{M_c \sin i_c}{a_c}^2 \sim
(0.121\pm0.014)\,\, M_{Jup}\,\, \textrm{AU}^{-2},
\end{equation}
where $i_c$ is the orbital inclination of ``c''
relative to the line of sight.
We plotted equation (7) (assuming $\sin i_c = 0$ for simplicity)
in figure~3 using a dashed-dotted line for reference.
Obviously, the Kozai migration scenario is totally excluded
if ``c'' existed in the outer region
(beyond the snow line) at the initial stage.
Thus in the presence of ``c'',
it is impossible to explain the tilted orbit of HAT-P-7b
by the Kozai migration caused by the distant binary companion only.

However, there is another chance of a ``sequential'' Kozai migration
scenario for a 2-body system with a binary star,
as introduced by \citet{2008ApJ...683.1063T}
and \citet{2010arXiv1003.0633K}.
Namely, an inclined binary companion induces the Kozai mechanism for
an outer body first, and then the inclined outer body leads
the Kozai mechanism for an inner body.
In this case, ``c'' could play
an important role for the migration of HAT-P-7b.
If this is the case, ``c'' could have a tilted orbital axis
relative to both the stellar spin axis and the orbital axis of HAT-P-7b.
We cannot discuss this possibility in detail at this point, because
the orbital parameters of ``c'' have not
yet been determined.
If orbital parameters of ``c''
are firmly determined, it would be interesting to discuss the 
possibility of a sequential Kozai migration scenario.
We also note that if the semi-major axis of ``c'' turns out to be large,
further direct imaging of this inner body would be also
interesting in the future (e.g., with TMT or E-ELT).

From the above discussions, we found that the Kozai migration scenario
caused by a distant binary star could be realized only
in a very limited situation.
In addition, if ``c'' existed, the Kozai migration
of HAT-P-7b caused directly by the binary star could not have occured,
although we could not refute the possibility of
a sequential Kozai migration for HAT-P-7b at this point.
In addition, if neither of the candidate stars
is a physical companion of HAT-P-7, only planet-planet
scattering can explain the tilted orbit of HAT-P-7b.
Thus with a few exceptions above, we conclude that planet-planet scattering
is a more plausible explanation for the migration mechanism of HAT-P-7b.

\section{Summary}

We conducted direct imaging observations of HAT-P-7 with
the Subaru HiCIAO and the Calar Alto AstraLux.
The system was reported to have the highly tilted transiting planet HAT-P-7b,
and massive bodies were expected to exist in the outer region
based on planetary migration theories.
We discovered two companion candidates around HAT-P-7.
We modeled and constrained the Kozai migration scenario for HAT-P-7b
under the existence of a binary star,
and found that the Kozai migration scenario was 
realizable only in a very limited condition and was not favored
if the additional body HAT-P-7c existed as reported by WJA09.
As a result, we conclude that planet-planet scattering is particularly
plausible for the migration mechanism of HAT-P-7b.
To complement our conclusion for the migration mechanism of HAT-P-7b,
further radial velocity measurements of HAT-P-7
are highly desired to constrain orbital parameters of HAT-P-7c.
In addition, further direct imaging and high spectral resolution
observations for the candidate companion stars would be very useful
to constrain common proper motions and peculiar radial velocities
of the stars and thereby to discriminate their binarity.

\bigskip

This paper is based on data collected at Subaru Telescope,
which is operated by the National Astronomical Observatory of Japan.
N.N. is supported by a Japan Society for Promotion of Science (JSPS)
Fellowship for Research (PD: 20-8141).
N.N, M.N., and S.I., were supported in part
by the National Science Foundation under Grant No. NSF PHY05-51164
(KITP program ``The Theory and Observation of Exoplanets'' at UCSB).
M.N. and B.S. are supported by
MEXT's program "Promotion of Environmental Improvement
for Independence of Young Researchers" under the special
Coordination Funds for Promoting Science and Technology.
M.N. acknowledges the support from MEXT's KAKENHI(21740324).
M.M. is supported by the U.S. National Science Foundation under
Award No. AST-0901967.
E.L.T. gratefully acknowledges the support from a Princeton University
Global Collaborative Research Fund grant and the World Premier
International Research Center Initiative (WPI Initiative), MEXT, Japan.
M.T. acknowledges support from The Mitsubishi Foundation.
We wish to acknowledge the very significant cultural role
and reverence that the summit of Mauna Kea has always had
for the people in Hawai'i.

%%%%%%%%%%%%%%%%%%%%%%%%%%%%%%%%%%%%%%%%%%%%%%%%%%%%%%%%%%%%%%%%%%%%%%

%%%%%%%%%%%%%%%%%%%%%%%%%%%%%%%%%%%%%%%%%%%%%%%%%%%%%%%%%%%%%%%%%%%%%%


\begin{thebibliography}{}
\expandafter\ifx\csname natexlab\endcsname\relax\def\natexlab#1{#1}\fi

\bibitem[{{Akeson} {et~al.}(1998){Akeson}, {Koerner}, \&
  {Jensen}}]{1998ApJ...505..358A}
{Akeson}, R.~L., {Koerner}, D.~W., \& {Jensen}, E.~L.~N. 1998, \apj, 505, 358

\bibitem[{{Baraffe} {et~al.}(2003){Baraffe}, {Chabrier}, {Barman}, {Allard}, \&
  {Hauschildt}}]{2003A&A...402..701B}
{Baraffe}, I., {Chabrier}, G., {Barman}, T.~S., {Allard}, F., \& {Hauschildt},
  P.~H. 2003, \aap, 402, 701

\bibitem[{{Borucki} {et~al.}(2009){Borucki}, {Koch}, {Jenkins}, {Sasselov},
  {Gilliland}, {Batalha}, {Latham}, {Caldwell}, {Basri}, {Brown},
  {Christensen-Dalsgaard}, {Cochran}, {DeVore}, {Dunham}, {Dupree}, {Gautier},
  {Geary}, {Gould}, {Howell}, {Kjeldsen}, {Lissauer}, {Marcy}, {Meibom},
  {Morrison}, \& {Tarter}}]{2009Sci...325..709B}
{Borucki}, W.~J., {et~al.} 2009, Science, 325, 709

\bibitem[{{Chatterjee} {et~al.}(2008){Chatterjee}, {Ford}, {Matsumura}, \&
  {Rasio}}]{2008ApJ...686..580C}
{Chatterjee}, S., {Ford}, E.~B., {Matsumura}, S., \& {Rasio}, F.~A. 2008, \apj,
  686, 580

\bibitem[{{Christensen-Dalsgaard} {et~al.}(2010){Christensen-Dalsgaard},
  {Kjeldsen}, {Brown}, {Gilliland}, {Arentoft}, {Frandsen}, {Quirion},
  {Borucki}, {Koch}, \& {Jenkins}}]{2010arXiv1001.0032C}
{Christensen-Dalsgaard}, J., {et~al.} 2010, ArXiv e-prints (CKB10)

\bibitem[{{Covey} {et~al.}(2007){Covey}, {Ivezi{\'c}}, {Schlegel},
  {Finkbeiner}, {Padmanabhan}, {Lupton}, {Ag{\"u}eros}, {Bochanski}, {Hawley},
  {West}, {Seth}, {Kimball}, {Gogarten}, {Claire}, {Haggard}, {Kaib},
  {Schneider}, \& {Sesar}}]{2007AJ....134.2398C}
{Covey}, K.~R., {et~al.} 2007, \aj, 134, 2398

\bibitem[{{Cutri} {et~al.}(2003){Cutri}, {Skrutskie}, {van Dyk}, {Beichman},
  {Carpenter}, {Chester}, {Cambresy}, {Evans}, {Fowler}, {Gizis}, {Howard},
  {Huchra}, {Jarrett}, {Kopan}, {Kirkpatrick}, {Light}, {Marsh}, {McCallon},
  {Schneider}, {Stiening}, {Sykes}, {Weinberg}, {Wheaton}, {Wheelock}, \&
  {Zacarias}}]{2003tmc..book.....C}
{Cutri}, R.~M., {et~al.} 2003, {2MASS All Sky Catalog of point sources.}, ed.
  {Cutri, R.~M., Skrutskie, M.~F., van Dyk, S., Beichman, C.~A., Carpenter,
  J.~M., Chester, T., Cambresy, L., Evans, T., Fowler, J., Gizis, J., Howard,
  E., Huchra, J., Jarrett, T., Kopan, E.~L., Kirkpatrick, J.~D., Light, R.~M.,
  Marsh, K.~A., McCallon, H., Schneider, S., Stiening, R., Sykes, M., Weinberg,
  M., Wheaton, W.~A., Wheelock, S., \& Zacarias, N.}

\bibitem[{{D'Angelo} {et~al.}(2002){D'Angelo}, {Henning}, \&
  {Kley}}]{2002A&A...385..647D}
{D'Angelo}, G., {Henning}, T., \& {Kley}, W. 2002, \aap, 385, 647

\bibitem[{{Daemgen} {et~al.}(2009){Daemgen}, {Hormuth}, {Brandner}, {Bergfors},
  {Janson}, {Hippler}, \& {Henning}}]{2009A&A...498..567D}
{Daemgen}, S., {Hormuth}, F., {Brandner}, W., {Bergfors}, C., {Janson}, M.,
  {Hippler}, S., \& {Henning}, T. 2009, \aap, 498, 567

\bibitem[{{Duch{\^e}ne} {et~al.}(2005){Duch{\^e}ne}, {Ghez}, {McCabe}, \&
  {Ceccarelli}}]{2005ApJ...628..832D}
{Duch{\^e}ne}, G., {Ghez}, A.~M., {McCabe}, C., \& {Ceccarelli}, C. 2005, \apj,
  628, 832

\bibitem[{{Duquennoy} \& {Mayor}(1991)}]{1991A&A...248..485D}
{Duquennoy}, A., \& {Mayor}, M. 1991, \aap, 248, 485

\bibitem[{{Fabrycky} \& {Tremaine}(2007)}]{2007ApJ...669.1298F}
{Fabrycky}, D., \& {Tremaine}, S. 2007, \apj, 669, 1298

\bibitem[{{Gaudi} \& {Winn}(2007)}]{2007ApJ...655..550G}
{Gaudi}, B.~S., \& {Winn}, J.~N. 2007, \apj, 655, 550

\bibitem[{{Hayano} {et~al.}(2008){Hayano}, {Takami}, {Guyon}, {Oya}, {Hattori},
  {Saito}, {Watanabe}, {Murakami}, {Minowa}, {Ito}, {Colley}, {Eldred},
  {Golota}, {Dinkins}, {Kashikawa}, \& {Iye}}]{2008SPIE.7015E..25H}
{Hayano}, Y., {et~al.} 2008, in Society of Photo-Optical
  Instrumentation Engineers (SPIE) Conference, 7015

\bibitem[{{Hioki} {et~al.}(2009){Hioki}, {Itoh}, {Oasa}, {Fukagawa}, {Kudo},
  {Mayama}, {Pyo}, {Hayashi}, {Hayashi}, {Ishii}, \&
  {Tamura}}]{2009PASJ...61.1271H}
{Hioki}, T., {et~al.} 2009, \pasj, 61, 1271

\bibitem[{{Hirano} {et~al.}(2010){Hirano}, {Suto}, {Taruya}, {Narita}, {Sato},
  {Johnson}, \& {Winn}}]{2010ApJ...709..458H}
{Hirano}, T., {Suto}, Y., {Taruya}, A., {Narita}, N., {Sato}, B., {Johnson},
  J.~A., \& {Winn}, J.~N. 2010, \apj, 709, 458

\bibitem[{{Hodapp} {et~al.}(2008){Hodapp}, {Suzuki}, {Tamura}, {Abe}, {Suto},
  {Kandori}, {Morino}, {Nishimura}, {Takami}, {Guyon}, {Jacobson},
  {Stahlberger}, {Yamada}, {Shelton}, {Hashimoto}, {Tavrov}, {Nishikawa},
  {Ukita}, {Izumiura}, {Hayashi}, {Nakajima}, {Yamada}, \&
  {Usuda}}]{2008SPIE.7014E..42H}
{Hodapp}, K.~W., {et~al.} 2008, in Society of Photo-Optical Instrumentation
  Engineers (SPIE) Conference, 7014

\bibitem[{{Hog} {et~al.}(1998){Hog}, {Kuzmin}, {Bastian}, {Fabricius},
  {Kuimov}, {Lindegren}, {Makarov}, \& {Roeser}}]{1998A&A...335L..65H}
{Hog}, E., {Kuzmin}, A., {Bastian}, U., {Fabricius}, C., {Kuimov}, K.,
  {Lindegren}, L., {Makarov}, V.~V., \& {Roeser}, S. 1998, \aap, 335, L65

\bibitem[{{Hormuth} {et~al.}(2008){Hormuth}, {Hippler}, {Brandner}, {Wagner},
  \& {Henning}}]{2008SPIE.7014E.138H}
{Hormuth}, F., {Hippler}, S., {Brandner}, W., {Wagner}, K., \& {Henning}, T.
  2008, in Society of Photo-Optical Instrumentation Engineers (SPIE) Conference
  Series, 7014

\bibitem[{{Ida} \& {Lin}(2004)}]{2004ApJ...616..567I}
{Ida}, S., \& {Lin}, D.~N.~C. 2004, \apj, 616, 567

\bibitem[{{Innanen} {et~al.}(1997){Innanen}, {Zheng}, {Mikkola}, \&
  {Valtonen}}]{1997AJ....113.1915I}
{Innanen}, K.~A., {Zheng}, J.~Q., {Mikkola}, S., \& {Valtonen}, M.~J. 1997,
  \aj, 113, 1915

\bibitem[{{Kita} {et~al.}(2010){Kita}, {Rasio}, \&
  {Takeda}}]{2010arXiv1003.0633K}
{Kita}, R., {Rasio}, F.~A., \& {Takeda}, G. 2010, ArXiv e-prints

\bibitem[{{K{\"o}hler} {et~al.}(2008){K{\"o}hler}, {Ratzka}, {Herbst}, \&
  {Kasper}}]{2008A&A...482..929K}
{K{\"o}hler}, R., {Ratzka}, T., {Herbst}, T.~M., \& {Kasper}, M. 2008, \aap,
  482, 929

\bibitem[{{Kozai}(1962)}]{1962AJ.....67..591K}
{Kozai}, Y. 1962, \aj, 67, 591

\bibitem[{{Kraus} \& {Hillenbrand}(2007)}]{2007AJ....134.2340K}
{Kraus}, A.~L., \& {Hillenbrand}, L.~A. 2007, \aj, 134, 2340

\bibitem[{{Lafreni{\`e}re} {et~al.}(2007){Lafreni{\`e}re}, {Marois}, {Doyon},
  {Nadeau}, \& {Artigau}}]{2007ApJ...660..770L}
{Lafreni{\`e}re}, D., {Marois}, C., {Doyon}, R., {Nadeau}, D., \& {Artigau},
  {\'E}. 2007, \apj, 660, 770

\bibitem[{{Lin} {et~al.}(1996){Lin}, {Bodenheimer}, \&
  {Richardson}}]{1996Natur.380..606L}
{Lin}, D.~N.~C., {Bodenheimer}, P., \& {Richardson}, D.~C. 1996, \nat, 380, 606

\bibitem[{{Lin} \& {Papaloizou}(1985)}]{1985prpl.conf..981L}
{Lin}, D.~N.~C., \& {Papaloizou}, J. 1985, in Protostars and Planets II, ed.
  D.~C. {Black} \& M.~S. {Matthews}, 981--1072

\bibitem[{{Marois} {et~al.}(2006){Marois}, {Lafreni{\`e}re}, {Doyon},
  {Macintosh}, \& {Nadeau}}]{2006ApJ...641..556M}
{Marois}, C., {Lafreni{\`e}re}, D., {Doyon}, R., {Macintosh}, B., \& {Nadeau},
  D. 2006, \apj, 641, 556

\bibitem[{{Marzari} \& {Weidenschilling}(2002)}]{2002Icar..156..570M}
{Marzari}, F., \& {Weidenschilling}, S.~J. 2002, Icarus, 156, 570

\bibitem[{{McLaughlin}(1924)}]{1924ApJ....60...22M}
{McLaughlin}, D.~B. 1924, \apj, 60, 22

\bibitem[{{Murray} \& {Dermott}(2000)}]{2000ssd..book.....M}
{Murray}, C.~D., \& {Dermott}, S.~F. 2000, {Solar System Dynamics} (Solar
  System Dynamics, by C.D.~Murray and S.F.~Dermott.~ ISBN 0521575974. UK:
  Cambridge University Press, 2000.)

\bibitem[{{Nagasawa} {et~al.}(2008){Nagasawa}, {Ida}, \&
  {Bessho}}]{2008ApJ...678..498N}
{Nagasawa}, M., {Ida}, S., \& {Bessho}, T. 2008, \apj, 678, 498

\bibitem[{{Narita} {et~al.}(2009){Narita}, {Sato}, {Hirano}, \&
  {Tamura}}]{2009PASJ...61L..35N}
{Narita}, N., {Sato}, B., {Hirano}, T., \& {Tamura}, M. 2009, \pasj, 61, L35
(NSH09)

\bibitem[{{Ohta} {et~al.}(2005){Ohta}, {Taruya}, \&
  {Suto}}]{2005ApJ...622.1118O}
{Ohta}, Y., {Taruya}, A., \& {Suto}, Y. 2005, \apj, 622, 1118

\bibitem[{{P{\'a}l} {et~al.}(2008){P{\'a}l}, {Bakos}, {Torres}, {Noyes},
  {Latham}, {Kov{\'a}cs}, {Marcy}, {Fischer}, {Butler}, {Sasselov}, {Sip{\H
  o}cz}, {Esquerdo}, {Kov{\'a}cs}, {Stefanik}, {L{\'a}z{\'a}r}, {Papp}, \&
  {S{\'a}ri}}]{2008ApJ...680.1450P}
{P{\'a}l}, A., {et~al.} 2008, \apj, 680, 1450 (PBT08)

\bibitem[{{Rasio} \& {Ford}(1996)}]{1996Sci...274..954R}
{Rasio}, F.~A., \& {Ford}, E.~B. 1996, Science, 274, 954

\bibitem[{{Rossiter}(1924)}]{1924ApJ....60...15R}
{Rossiter}, R.~A. 1924, \apj, 60, 15

\bibitem[{{Suzuki} {et~al.}(2010)}]{2010Suzuki}
{Suzuki}, R., {et~al.} 2010, in prep.

\bibitem[{{Takeda} {et~al.}(2008){Takeda}, {Kita}, \&
  {Rasio}}]{2008ApJ...683.1063T}
{Takeda}, G., {Kita}, R., \& {Rasio}, F.~A. 2008, \apj, 683, 1063

\bibitem[{{Takeda} \& {Rasio}(2005)}]{2005ApJ...627.1001T}
{Takeda}, G., \& {Rasio}, F.~A. 2005, \apj, 627, 1001

\bibitem[{{Tamura} {et~al.}(2006){Tamura}, {Hodapp}, {Takami}, {Abe}, {Suto},
  {Guyon}, {Jacobson}, {Kandori}, {Morino}, {Murakami}, {Stahlberger},
  {Suzuki}, {Tavrov}, {Yamada}, {Nishikawa}, {Ukita}, {Hashimoto}, {Izumiura},
  {Hayashi}, {Nakajima}, \& {Nishimura}}]{2006SPIE.6269E..28T}
{Tamura}, M., {et~al.} 2006, in Society of Photo-Optical
  Instrumentation Engineers (SPIE) Conference, 6269

\bibitem[{{Thalmann} {et~al.}(2009){Thalmann}, {Carson}, {Janson}, {Goto},
  {McElwain}, {Egner}, {Feldt}, {Hashimoto}, {Hayano}, {Henning}, {Hodapp},
  {Kandori}, {Klahr}, {Kudo}, {Kusakabe}, {Mordasini}, {Morino}, {Suto},
  {Suzuki}, \& {Tamura}}]{2009ApJ...707L.123T}
{Thalmann}, C., {et~al.} 2009, \apjl, 707, L123

\bibitem[{{Welsh} {et~al.}(2010){Welsh}, {Orosz}, {Seager}, {Fortney},
  {Jenkins}, {Rowe}, {Koch}, \& {Borucki}}]{2010arXiv1001.0413W}
{Welsh}, W.~F., {Orosz}, J.~A., {Seager}, S., {Fortney}, J.~J., {Jenkins}, J.,
  {Rowe}, J.~F., {Koch}, D., \& {Borucki}, W.~J. 2010, ArXiv e-prints (WOS10)

\bibitem[{{Winn} {et~al.}(2009){Winn}, {Johnson}, {Albrecht}, {Howard},
  {Marcy}, {Crossfield}, \& {Holman}}]{2009ApJ...703L..99W}
{Winn}, J.~N., {Johnson}, J.~A., {Albrecht}, S., {Howard}, A.~W., {Marcy},
  G.~W., {Crossfield}, I.~J., \& {Holman}, M.~J. 2009, \apjl, 703, L99 (WJA09)

\bibitem[{{Wu} \& {Murray}(2003)}]{2003ApJ...589..605W}
{Wu}, Y., \& {Murray}, N. 2003, \apj, 589, 605

\bibitem[{{Wu} {et~al.}(2007){Wu}, {Murray}, \&
  {Ramsahai}}]{2007ApJ...670..820W}
{Wu}, Y., {Murray}, N.~W., \& {Ramsahai}, J.~M. 2007, \apj, 670, 820

\end{thebibliography}
\end{document}